\documentclass[osajnl,superscriptaddress,11pt,preprint,a4paper,floatfix]{revtex4-1}
\usepackage{amsmath}
\usepackage{amsfonts}
\usepackage{amssymb}
\usepackage{amsthm}
\usepackage{xspace}

\usepackage{graphicx}
\usepackage{xcolor}
\usepackage{tikz}
\usetikzlibrary{external}
\usepackage{array}
\usepackage{booktabs}

\usepackage[sfdefault=cmbr]{isomath}
\usepackage{siunitx}

\usepackage{myacro}
\usepackage{mycom}
\newcommand{\blue}{\SI[detect-all]{399}{nm}\xspace}
\newcommand{\green}{\SI[detect-all]{556}{nm}\xspace}

\usepackage[final]{hyperref}
\usepackage[all]{hypcap}
\usepackage{cleveref}

\usepackage[grapics,tightpage]{preview}
\PreviewEnvironment{picture}

\crefname{figure}{Fig.}{Figs.}
\Crefname{figure}{Figure}{Figures}
\crefname{subfigure}{Fig.}{Figs.}
\Crefname{subfigure}{Figure}{Figures}
\crefname{equation}{Eq.}{Eqs.}
\Crefname{equation}{Equation}{Equations}

\hypersetup{
pdfauthor = {Marco Pizzocaro},
colorlinks,%
citecolor=black,%
filecolor=black,%
linkcolor=black,%
urlcolor=blue
}

\newcommand{\inputgnuplot}[1]{\includegraphics{#1-full}}
\begin{document}

%\preprint{}

%\title{\SI[detect-all]{80}{\%} Efficiency Frequency Doubling at \blue} %Title of paper
\title{Efficient Frequency Doubling at \blue}
%\title{Generation of a \SI[detect-all]{1}{W} laser at \blue by \acl{shg}} %Title of paper



\author{Marco Pizzocaro}
\email[Author to whom correspondence should be addressed. Electronic mail: ]{m.pizzocaro@inrim.it}
%\thanks{}
%\altaffiliation{f. INRIM}
\affiliation{Politecnico di Torino, Dipartimento di Elettronica e Telecomunicazioni, C.so duca degli Abruzzi 24, 10125 Torino, Italy}
\affiliation{Istituto Nazionale di Ricerca Metrologica (INRIM), Str. delle Cacce 91, 10135 Torino, Italy}

\author{Davide Calonico}
\affiliation{Istituto Nazionale di Ricerca Metrologica (INRIM), Str. delle Cacce 91, 10135 Torino, Italy}

\author{Pablo Cancio Pastor}
\affiliation{Istituto Nazionale di Ottica (INO-CNR), Via Nello Carrara, 1, 50019 Sesto Fiorentino, Italy}
\affiliation{European Laboratory for Non-Linear Spectroscopy (LENS), Via Nello Carrara, 1, 50019 Sesto Fiorentino, Italy}

\author{Jacopo~Catani}
\affiliation{Istituto Nazionale di Ottica (INO-CNR), Via Nello Carrara, 1, 50019 Sesto Fiorentino, Italy}
\affiliation{European Laboratory for Non-Linear Spectroscopy (LENS), Via Nello Carrara, 1, 50019 Sesto Fiorentino, Italy}

\author{Giovanni A. Costanzo}
\affiliation{Politecnico di Torino, Dipartimento di Elettronica e Telecomunicazioni, C.so duca degli Abruzzi 24, 10125 Torino, Italy}

\author{Filippo Levi}
\affiliation{Istituto Nazionale di Ricerca Metrologica (INRIM), Str. delle Cacce 91, 10135 Torino, Italy}

\author{Luca Lorini}
\affiliation{Istituto Nazionale di Ricerca Metrologica (INRIM), Str. delle Cacce 91, 10135 Torino, Italy}

%\author{Etc. Etc.}
%\affiliation{Politecnico di Torino, Dipartimento di Elettronica e Telecomunicazioni, C.so duca degli Abruzzi 24, 10125 Torino, Italy}
%\affiliation{Istituto Nazionale di Ricerca Metrologica (INRIM), Str. delle Cacce 91, 10135 Torino, Italy}


\date{\today}

\begin{abstract}
This article describes a reliable, high-power, and narrow-linewidth laser source at \blue useful for cooling and trapping of ytterbium atoms.
A continuous-wave \acl{tisa} laser at \SI{798}{nm} is frequency doubled using a \acl{lbo} crystal in an enhancement cavity.
Up to \SI{1.0}{W} of light at \blue has been obtained from \SI{1.3}{W} of infrared light, with an efficiency of \SI{80}{\%}.

\end{abstract}

%A high-power and narrow-linewidth laser source at 399 nm is obtained by frequency doubling of a continuous-wave titanium-sapphire laser at 798 nm.  Up to 1.0 W of light at 399 nm has been obtained from 1.3 W of infrared light, with an efficiency of 80%, using a lithium triborate crystal in an enhancement cavity.  This source is useful for for cooling and trapping of ytterbium atoms. 


\maketitle %\maketitle must follow title, authors, abstract and \pacs

Ytterbium holds interest for several atomic physics experiments because it is easy to cool and trap and has seven stable isotopes.
Such experiments include optical frequency standards \citep{Hinkley2013}, non-conservation measurement \citep{DeMille1995}, Bose-Einstein condensation \citep{Takasu2003}, degenerate Fermi gases \citep{Fukuhara2007,Pagano2014}, bosonic-fermionic systems \citep{Honda2002}, and quantum information \citep{Hayes2007}.
Ytterbium atoms can be cooled at millikelvin temperature using a \mot on the strong blue transition \ce{^1S0 - ^1P1} at \blue.
Microkelvin temperatures can be achieved with a \mot on the narrower intercombination transition \ce{^1S0 - ^3P1} at \green.
In practical cases the \green \mot needs an efficient loading stage that uses \blue radiation.
The blue \ce{^1S0 - ^1P1} transition has a saturation intensity of \SI{60}{mW/cm^2} and while few milliwatts are enough to trap atoms in a \blue \mot \cite{Park2003}, the laser radiation power limits the full exploitation of the cooling process until all beams reach the saturation.
In experiments needing fast loading times (\eg optical clocks) the \green \mot is loaded starting from the \blue \mot, where larger laser beams (typically with a diameter of \SI{1}{cm}) increase the rate of capture in the trap and the number of trappable atoms.
In a degenerate gases experiment looking  for maximum atomic density the \green \mot can be directly loaded exploiting a Zeeman slower based on the \blue transition, which usually requires hundreds of milliwatts of \blue light.
So, in both cases, a reliable and powerful laser source at \blue is important for efficient trapping and cooling of ytterbium.

In this paper we present a high-power and narrow-linewidth laser source at \blue used for a ytterbium optical frequency standard experiment developed at \inrim.
%This wavelength can be reached using \ac{ingan} diode lasers \citep{Park2003}.
%However, the availability of laser diode at this wavelength is limited because manufacturers concentrates on the close \SI{405}{nm} wavelength of Blu-ray Discs.
Even if \ac{ingan} diode lasers can emit at \blue \citep{Park2003} their availability at this wavelength is limited because manufacturers concentrates on the close \SI{405}{nm} wavelength of Blu-ray Discs.
As well, their power is typically less than \SI{50}{mW}.
The best choice to get a power boosted coherent source at such wavelength is the \shg in a nonlinear crystal.
Some \shg sources are commercially available, but with limited power.
%The blue \blue radiation comes from \shg of a \SI{798}{nm} \tisa laser in a \ac{lbo} crystal, chosen to achieve high power, stability and tunability \citep{Adams1992,Kumagai2003,Scheid2007,Cha2010,Ferrari2010}.
% at \blue of a \SI{798}{nm}
The \shg of a \tisa laser in a \ac{lbo} crystal achieves high power, stability and tunability \citep{Adams1992,Kumagai2003,Koelemeij2005,Scheid2007,Cha2010,Ferrari2010}.
%The non-linear crystal is a bulk \ac{lbo} crystal and phasematching is achieved by angle-tuning (critical phase matching).
%\ac{lbo} has been preferred to \ac{ppktp} because it is transparent from \SI{160}{nm} to \SI{2600}{nm} and thus does not have thermal effects from absorption at \SI{399}{nm} that we observed with \ac{ppktp}.
%The availability of laser diode at this wavelength is limited because manufacturers concentrates on the close \SI{405}{nm} wavelength of Blu-ray Disc.
Another popular choice for the \shg of blue light is \ac{ppktp} that can achieve high conversion efficiency \citep{Targat2005}.
Even if \ac{ppktp} is better suited for a low-power blue source, since it is more efficient, the wavelength \blue is at the edge of its transparency range (\SIrange{350}{4400}{nm}) and photo-refractive effects at \blue make it unsuitable for \shg at high power.
%We observed that the absortion of \ac{ppktp} at \blue was limiting the performances of \shg at this wavelength.
\ac{lbo} is transparent from \SI{160}{nm} to \SI{2600}{nm} and initial studies performed at the \ac{ino} and \ac{lens} showed that it is free of thermal effects from absorption at \SI{399}{nm}. 
We show that \SI{1.0}{W} of blue light at \blue can be generated from \SI{1.3}{W} of infrared light using a \ac{lbo} crystal in an enhancement cavity.

\section{Second harmonic generation with enhancement cavity}

%\ac{lbo} crystal does not allow temperature-tuned phase-matching for the \shg at \blue, and it must be angle tuned (type I critical phase-matching) taking into account the walk-off angle $\rho$ between the fundamental and duplicated waves.
\ac{lbo} crystals allow only birefringent phase-matching for \shg because they are not ferroelectric.
Both crystal temperature and orientation tuning can be performed for phase matching.
%The required high temperature values at these wavelengths make preferable angle-tuned phase-matching for practical issues.
Angle-tuned phase-matching is preferable in practice because of the high temperature values required for temperature tuning at these wavelengths.  
We perform type I critical phase-matching taking into account the walk-off angle $\rho$ between the fundamental and frequency-doubled waves.

The second harmonic power is $P_3 = E\ped{nl} P_1^2$, where $P_1$ is the pump power on the crystal at the fundamental wavelength.
The non-linear coefficient $E\ped{nl}$ can be calculated for Gaussian beams by the Boyd and Kleinman formula
 \citep{Boyd1968} that in the International System (SI) units is \citep{Risk2003}
\begin{equation}\label{eq:boyd}
E\ped{nl} = \frac{16 \pi^2 d\ped{eff}^2 l}{\epsilon_0 c \lambda_1^3 n_3 n_1} \eu^{-\alpha' l} h\ped{m}(B, \xi),
\end{equation}
where $d\ped{eff}$ is the effective non-linear coefficient of the crystal for \shg, $l$ is the length of the crystal, $n_1$ and $n_3$ are the index of refraction for the fundamental and second harmonic frequencies respectively, and $\alpha' = \alpha_1+\alpha_3/2$ accounts for absorption of the fundamental ($\alpha_1$) and of the second harmonic ($\alpha_3$).
The function $h\ped{m}(B, \xi)$ (Boyd-Kleinman factor) accounts for the walk-off angle $\rho$  ($B=\rho \sqrt{\pi l n_1/2\lambda_1}$) and the focusing of the Gaussian beam $\xi=l/2z\ped{R}$, where $z\ped{R}=\pi w_0^2/\lambda_1$ is the Rayleigh range of the Gaussian mode of the fundamental wavelength $\lambda_1$ and $w_0$ is its beam waist radius.  
Depending on $B$, $h\ped{m}$ has a maximum for $\xi$ between 1.39 and 2.64.
%For $B=0$ (no walk-off) the maximum of  $h\ped{m}\simeq1$.
%Instead $h\ped{m}$ strongly reduce the \shg power for large $B$. 
%Some properties of \ac{lbo} are reported in \cref{tab:lbo}.
\Cref{tab:lbo} summarizes some properties of \ac{lbo} for the \shg at \blue \citep{Smith2009}.
We estimated $\alpha_3\approx\SI{3.1e-1}{m^{-1}}$ (from the coefficient between \SIrange{351}{364}{nm}) and $\alpha_1\approx\SI{3.5e-2}{m^{-1}}$ (from the coefficient at \SI{1064}{nm}).
Calculations lead to $B=3.5$ and the Boyd-Kleinman factor of \cref{eq:boyd} has a maximum of $h\ped{m}=0.19$ for  $\xi=1.47$ \citep{Boyd1968}. For a crystal length $l=\SI{15}{mm}$ this corresponds to an optimal waist radius $w_0=\sqrt{z\ped{R}\lambda_1/\pi}=\SI{36}{\micro m}$. %a Rayleigh range $z\ped{R}=\SI{4.8}{mm}$ or
The uncertainty in the estimate of $E\ped{nl}$ is dominated by $d\ped{eff}$, where we assume a relative uncertainty of $\SI{7}{\%}$ from the properties of the crystal \cite{Smith2009,Velsko1991}.
Then \cref{eq:boyd} predicts a non-linear coefficient $E\ped{nl}=\SI{7(1)e-5}{W^{-1}}$.

\begin{table}
\caption{\label{tab:lbo}Properties of \ac{lbo} crystal for \shg from $\SI{798}{nm}$ to \blue.}
\begin{ruledtabular}
\begin{tabular}{lll}

Parameter				&	&				\\
\hline
Nonlinear coefficient	& $d\ped{eff}$	&	\SI{0.75}{pm/V}\\
Cutting angles & $\theta$ & \SI{90}{\degree}\\
			 & $\phi$ & \SI{31.8}{\degree}\\
Walk-off angle	& $\rho$	& \SI{0.0162}{rad}\\
Ord. index of refr. at \SI{798}{nm} & $n\ped{o}(\omega_1)$ & \num{1.611}\\
Extr. index of refr. at \SI{399}{nm} & $n\ped{e}(\omega_3)$ & \num{1.611}\\
%Absortion at \SI{798}{nm} &	$\alpha_1$ & \SI{3.5e-2}{m^{-1}}\\
%Absortion at \SI{399}{nm} &	$\alpha_3$ & \SI{3.1e-1}{m^{-1}}\\

%	deff = 0.75e-12		
%	ro = 0.0162		

%	ab1=3.5e-2		
%	ab3=3.1e-1	

\end{tabular}
\end{ruledtabular}
\end{table}

The power circulating in the enhancing cavity  at the fundamental wavelength $P\ped{c}$ depends on the power injected in the cavity $P\ped{in}$ as \cite{Polzik1991} %injected
\begin{equation}\label{eq:pic}
P\ped{c} = \frac{T_1 P\ped{in} }{\left[1-\sqrt{(1-T_1) (1-l\ped{cav}) (1-E\ped{nl} P\ped{c})}\right]^2},
\end{equation}
where $T_1$ is the transmission of the input coupler of the cavity and $l\ped{cav}$ is the linear loss in the cavity. % and $\eta\ped{m}$. %, by the crystal and mirrors other than the input coupler and $E\ped{nl}=\SI{7.5e-5}{W^{-1}}$ is the effective nonlinear coefficient for \ac{lbo} \cite{Boyd1968}.
Here we assume a  mode-matched cavity, otherwise $P\ped{in}$ has to be replaced with the the power of light coupled in the cavity $P\ped{in} \rightarrow P\ped{ic} = \eta P\ped{in}$, where $\eta$ is the mode-matching efficiency. % in \cref
The \shg power is then $P\ped{out} = E\ped{nl} P\ped{c}^2$.
%The optimal input coupler transmission depends on the linear losses and the quadratic losses introduced by the \shg
From \cref{eq:pic} the circulating power $P\ped{c}$ is maximized choosing an input coupler with transmission
\begin{equation}\label{eq:topt}
T\ped{opt}=\frac{l\ped{cav}}{2}+\sqrt{\frac{l\ped{cav}^2}{4}+E\ped{nl} P\ped{in}}.
\end{equation}
Choosing $T_1=T\ped{opt}$ corresponds to an impedance-matched cavity with zero reflection on resonance.

We initially estimated  $l\ped{cav}\approx\num{4e-3}$ from the specifications of crystal and mirrors coatings.
Assuming $P\ped{in} = \SI{1.2}{W}$ we choose the input coupler with a transmission $T_1 = \SI{1.2}{\%}$.

%If the quadratic depletion of the pump is negligible ($(1-E\ped{nl} P\ped{c}) \approx 1$), the  power circulating in the cavity is maximized choosing an input coupler transmission $T_1=l\ped{cav}$.
%This situation is a sort of impedance matching in which the power injected in the cavity corresponds exactly to the round trip power loss.
%If the depletion of the pump is not negligible, that is for high efficient \shg, \cref{eq:pic} should in general be solved numerically for $P\ped{c}$.

%where $E\ped{nl}=\SI{7.5e-5}{W^{-1}}$ is calculated from \cref{eq:boyd}.

\section{Experimental setup}
%In our experiment the blue \blue radiation is obtained by \shg from a \SI{798}{nm} \tisa laser using a \ac{lbo} crystal, chosen to achieve high power, stability and tunability \cite{Ferrari2010}.

The \SI{798}{nm} source is a \tisa laser pumped by a \SI{8}{W} solid state pump laser at \SI{532}{nm}.
It can be tuned from \SIrange{700}{970}{nm} and it  has an output power of \SI{1.3}{W}  at \SI{798}{nm} with a linewidth $<\SI{20}{kHz}$. 

Our \ac{lbo} crystal is $l=\SI{15}{mm}$ long, has a section of \SI{3x3}{mm} and it is made by Raicol.
It is cut for normal incidence at the phase-matching angle (cutting angles $\theta=\SI{90}{\degree}$ and $\phi=\SI{31.8}{\degree}$) and has anti-reflection coating on the faces for both wavelengths, $R(\SI{798}{nm})<\SI{0.1}{\%}$ and $R(\SI{399}{nm})<\SI{0.3}{\%}$.

\begin{figure}
\centering
\includegraphics{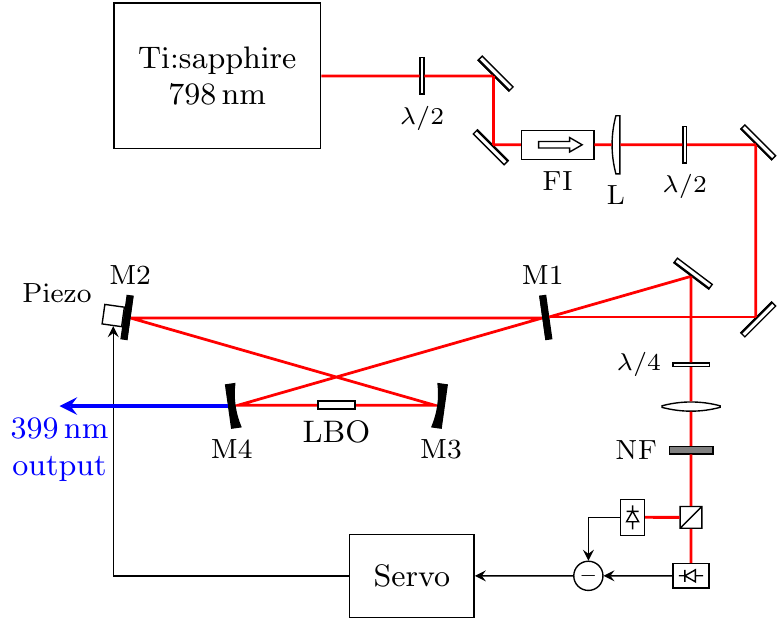}
\caption{Block scheme of the \shg setup. FI: Faraday Isolator. L: mode matching lens. M1--M4: cavity mirrors. NF: neutral density filter.}
\label{fig:block}
\end{figure}

To increase the output power at \blue the crystal is placed in a bow-tie enhancement cavity, resonant at \SI{798}{nm}. % (see the cavity sketch in \cref{fig:bluecav}).
The optical setup is sketched in \cref{fig:block}.
The \ac{lbo} crystal is held by a copper mounting on a rotational stage.
The crystal works at room temperature and its temperature is not actively stabilized.
The two mirrors close to the crystal (M3 and M4 in the figure) are concave with a radius of curvature $r=\SI{100}{mm}$.
The input coupler M1 and mirror M2 are flat.
The total length of the cavity is \SI{715}{mm}, the \fsr is \SI{420}{MHz}, and the distance between the curved mirrors is \SI{114}{mm}.
The angle of incidence of the beam on the mirrors (folding angle) is $\alpha=\SI{8}{\degree}$. It is as small as possible to limit the astigmatism caused by the tilted curved mirrors.
This geometry leads to a circular beam waist inside the crystal of \SI{35}{\micro m} as calculated by $ABCD$ matrix formalism \cite{Freegarde2001} that is close to the optimal value for the Boyd-Kleinman factor.
%This value is close to the optimum value for \shg \cite{Boyd1968}.

The secondary waist between M1 and M2 is slightly elliptical, the waist radius in the horizontal and vertical direction are calculated to be $w_{2x} = \SI{261}{\micro m}$ and $w_{2y} = \SI{295}{\micro m}$. % (ellipticity = wy/wx = 0.885).
The light from the \tisa laser is coupled to this waist using a single mode-matching lens (focal length $f=\SI{400}{mm}$).
A Faraday isolator prevents reflections on the laser.
The input polarization is vertical. The resulting output at \blue is horizontally polarized.

\begin{figure}
\centering
\includegraphics[width=0.5\textwidth]{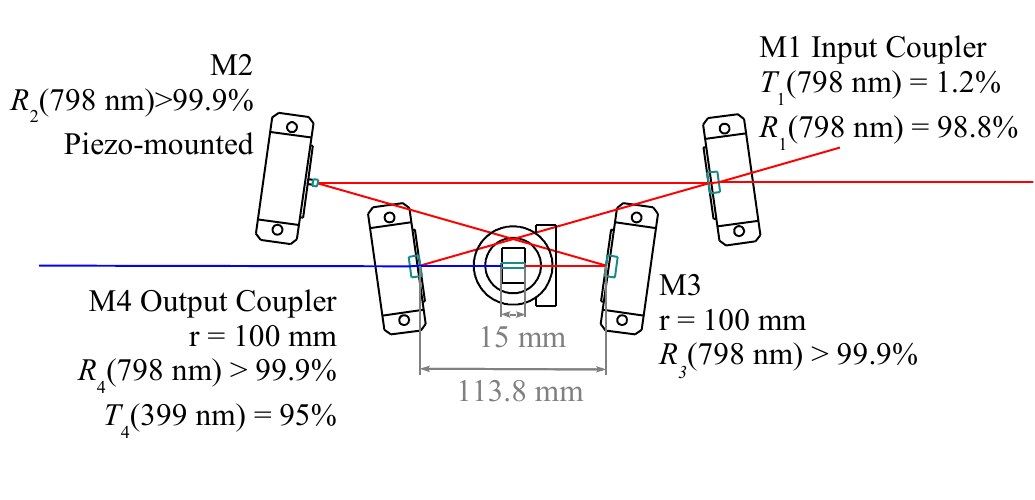}\\
\includegraphics[width=0.4\textwidth]{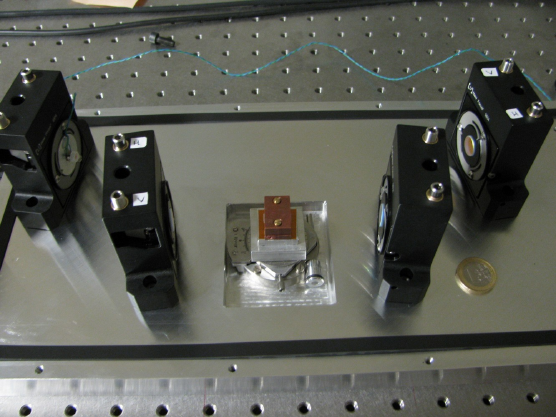}
\caption{Draw and picture of the cavity for the \shg of \blue with the \ac{lbo} crystal.}
\label{fig:bluecav}
\end{figure}

The nominal reflectivity of the input coupler M1 is $R_1 = \SI{98.8}{\%}$ and the nominal transmission is $T_1 = \SI{1.2}{\%}$.
Mirrors M2 and M3 are high-reflection coated at \SI{798}{nm} ($R_2$ and $R_3>\SI{99.9}{\%}$).
The output coupler M4 is coated for high-reflection at \SI{798}{nm} and anti-reflection at \blue ($T_4(\blue)=0.95$).

For maximum stability and simplicity in the alignment we mounted the mirrors on top-actuated mounts placed on a monolithic block of aluminum (see \cref{fig:bluecav}). 
The blue light assists chemical reactions in dust particles that may damage the crystal surfaces.
Moreover \ac{lbo} crystals are reported to be degraded by humidity \citep{Cha2010}.
Studies of the cavity performances done at \ac{ino} and \ac{lens}  has shown the importance of air-tight environment in
order to achieve long life of the crystal.
In a similar setup at \ac{ino} and \ac{lens} the entire cavity is under vacuum, and it has been operated for more than two years without degradation (pump input power $\SI{1}{W}$, output power \SI{0.6}{W}).
At \inrim the cavity is closed in an air-tight aluminum box to prevent contamination of the optics by dust, but for simplicity it is not under vacuum.
Silica-gel is put in the box to reduce humidity even if we did not observed degradation from the \SI{50}{\%} relative humidity in the lab. % that has been reported to degrade the crystal faces\citep{Cha2010}.
We have observed no degradation of the \shg for the last year of operations under these conditions.

The flat mirror M2 is mounted on a small piezo-electric actuator (\SI{3 x 3}{mm}) to lock the cavity on the \tisa laser with the \hc technique \citep{Hansch1980}.
The piezo spans about 6 free spectral ranges of the cavity (\SI{\sim2.5}{GHz} of the fundamental).
In the \hc technique the \ac{lbo} crystal is the polarizing element inside the cavity. 
The half-waveplate in front of the cavity is rotated to give a small angle to the polarization of the incident radiation with respect to axis of the crystal.
The reflection from the input coupler is attenuated by a beam-sampler and a neutral density filter and the \hc signal is obtained by a polarization analyzer (quarter-waveplate and polarizing cube).
A servo keeps the cavity on resonance with the laser frequency with a bandwidth of \SI{10}{kHz}.
We observed that a reliable lock requires good optical isolation of the pump (\SI{>30}{dB}), whether the pump is a \tisa or an amplified diode laser.

%For example the dependence of the \shg power with the transmission of the input coupler M1 for different linear losses is shown in \cref{fig:ic}.
%The input coupler transmission should then be chosen to match the input power and the linear losses of the actual cavity.
%We estimated a linear loss in the cavity coming from the mirrors and the coating on the faces of the crystal $l\ped{cav} \sim \num{4e-3}$.
%For a well matched input power of \SI{1.1}{W} the transmission of the input coupler was chosen $T_1 = \num{0.012}$

\section{Results}

\begin{figure}
\centering
\inputgnuplot{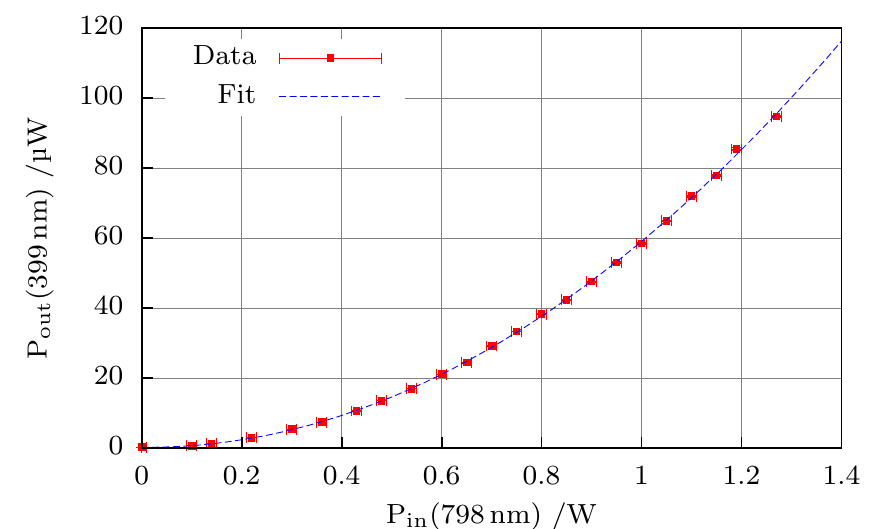}
\caption{\shg power from the \ac{lbo} crystal used in single pass, at the output of the crystal.}
\label{fig:single}
\end{figure}

We measured all the relevant properties for the conversion in the cavity.
The single-pass conversion was measured replacing the input coupler M1 with an anti-reflection coated window.
From a quadratic fit of the output power at \blue we obtained a nonlinear conversion efficiency $E\ped{nl} = \SI{6.0(3)e-5}{W^{-1}}$ for the power just outside the crystal surface (see \cref{fig:single}).
This value is consistent with the expected one within their uncertainties.
%This value is \SI{83}{\%} of the theoretical value.
%The discrepancy from the expected single pass conversion and the measured one arises from uncertainty in the properties of the crystal, that are given at \SI{1064}{nm}, and from the uncertainty in the properties of the gaussian beam used for calculating the Boyd and Kleinmann factor $h\ped{m}$.
%Kumagai et al. measured a similar ratio for the \shg at \SI{371}{nm} \citep{Kumagai2003}.
The linear losses in the cavity were directly measured  from the finesse after replacing the input coupler M1 with a high-reflection coated mirror.
%The finesse resulted $\mathcal{F} = 6510(30)$ and the linear losses $l\ped{cav}=\num{9.65(4)e-4}$.
The finesse resulted $\mathcal{F} = 6500(100)$ and the linear losses $l\ped{cav}=\num{9.7(1)e-4}$.

The transmission of the input coupler M1 is $T_1=\SI{1.32(2)}{\%}$.
From \cref{eq:topt}, the optimal input coupler transmission is $T\ped{opt}=\SI{0.93}{\%}$ at $P\ped{in} = \SI{1.2}{W}$ using the measured values of $E\ped{nl}$, $l\ped{cav}$.
Since $T_1 \neq T\ped{opt}$ the cavity is not exactly impedance matched.
%From the measured values of $E\ped{nl}$ and $l\ped{cav}$ the optimal input coupler transmission is $T\ped{opt}=\SI{0.93}{\%}$, as calculated from \cref{eq:topt}.
The contrast of the fringe in reflection is $\SI{93(1)}{\%}$ for $P\ped{in} = \SI{1.2}{W}$. 
This value is consistent with a mode matching efficiency of $\eta = \SI{98(1)}{\%}$ since we expect a contrast of $\SI{95.5(5)}{\%}$ from impedance mismatch \cite{Koelemeij2005}. %:
%\begin{equation}
%reflected power on resonance = \frac{\left( \sqrt{1-T_1} -  \sqrt{(1-l\ped{cav}) (1-E\ped{nl} P\ped{c})} \right)^2}{\left[1-\sqrt{(1-T_1) (1-l\ped{cav}) (1-E\ped{nl} P\ped{c})}\right]^2}
%\end{equation}

%The optimal input coupler transmission is $T\ped{opt}=\SI{0.93}{\%}$, as calculated from \cref{eq:topt} using the measured values of $E\ped{nl}$, $l\ped{cav}$, and $\eta$.

\begin{figure}
\centering
\inputgnuplot{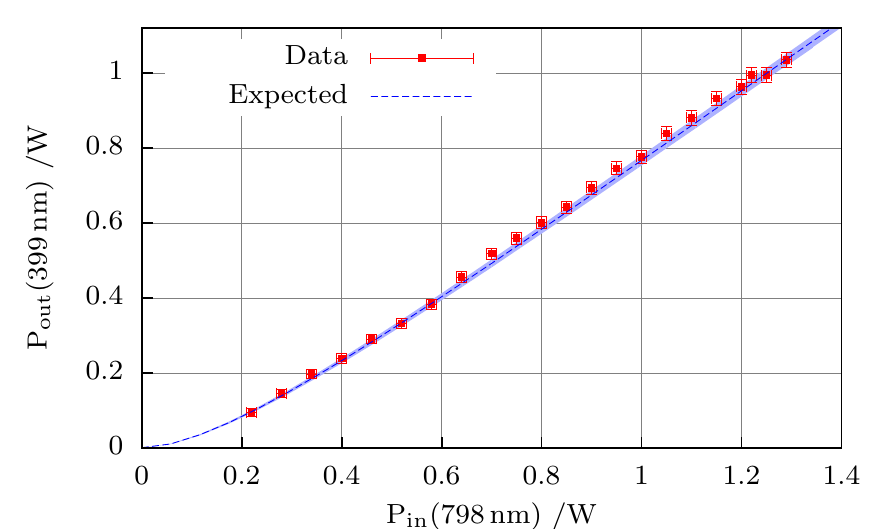}
\caption{\shg power out of the cavity as a function of input power. Last two data points are obtained increasing the power of the \tisa pump above the normal value of \SI{8.0}{W}. %The line is theoretically calculated for $E\ped{nl} = \SI{6.0e-5}{W^{-1}}$, $T_1=\num{1.32e-2}$ and $l\ped{cav}=\num{9.7e-4}$.
The dashed line shows the theoretical expectation from the measured values of $E\ped{nl}$, $l\ped{cav}$, $T_1$, and $\eta$.
Shaded region denotes uncertainty from the parameters.
}
\label{fig:shg800}
\end{figure}

\Cref{fig:shg800} shows experimental data for the \shg power at \blue as a function of the input power.
The blue output was measured after two lenses and a window considering a total propagation transmission of \SI{94}{\%}.
A narrowband blue filter was used to stop the leaking infrared light  (filter transmission $\SI{42}{\%}$ at \blue and  $<\num{1e-5}$ at \SI{798}{nm}).
Without filter the power of the leaking infrared light is about \SI{3}{\%} of the measure of the output power (after 3 optical elements coated for the blue).
The output is $P\ped{out} = \SI{0.99(2)}{W}$ with an input power $P\ped{in}=\SI{1.21(1)}{W}$.
Increasing the power of the \tisa pump from \SI{8.0}{W} to \SI{8.5}{W} allowed for an input power of $P\ped{in}=\SI{1.29(1)}{W}$ and an output of $P\ped{out}=\SI{1.04(2)}{W}$.
The maximum observed efficiency is $\epsilon = P\ped{out}/P\ped{in} = \SI{81(2)}{\%}$ for $P\ped{in} = \SI{1.22}{W}$.

The same figure shows also the theoretical prediction from the measured values of the nonlinear conversion efficiency, cavity linear losses, input coupler transmission, and mode-matching efficiency.
A numerical, self-consistent algorithm has been used to compute the power circulating in the cavity as in \cref{eq:pic} and then the output power.
%From the measured values of $E\ped{nl}$ and $l\ped{cav}$ the optimal input coupler transmission is $T\ped{opt}=\SI{0.93}{\%}$, as calculated from \cref{eq:topt}.
%%The same theory predicts that replacing our input coupler with the optimal one should result in a relative increase of the output power for $P\ped{in}=\SI{1.2}{W}$ below \SI{3}{\%}.
%The mismatch between the optimal  input coupler  and the actual one ($T_1=\SI{1.32(2)}{\%}$) is not critical, since  for $P\ped{in}=\SI{1.2}{W}$ the theory predicts a decrease in output power below \SI{3}{\%}.

%The blue beam is elliptical and we use two cylindrical lenses to make it circular and without astigmatism.
The walk-off angle $\rho$ and the bow-tie cavity make the blue beam astigmatic and elliptical.
The \blue beam is reshaped by two cylindrical lenses and then it is collimated by a single spherical lens.% and separated in two beams that are frequency-shifted by \aclp{aom}.
After frequency-shifting using \aclp{aom} we coupled the blue light in two \acl{pm} fibers with \SI{3.5}{\micro m} core.
We achieved a coupling efficiency up to \SI{72}{\%} using two lenses for mode-matching.

%Theory and experimental data are well matched; they slightly diverge at high power because each data point was taken without re-alignment of the optics. 
%This system replaced a similar one that used a \ac{ppktp} crystal that was found to be too absorptive at \blue and was delivering a maximum output power of \SI{60}{mW}.

\begin{figure}
\centering
\inputgnuplot{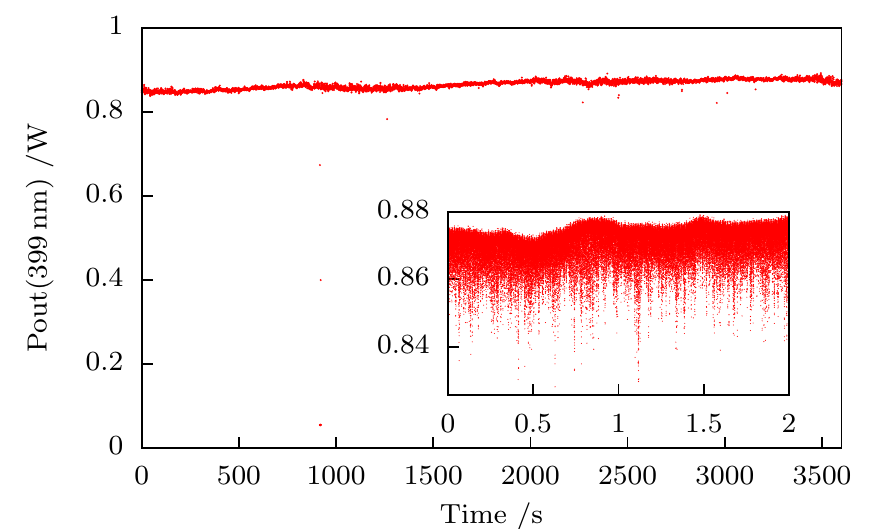}
\caption{Power output of the \shg as a function of time. Inset shows a detail at short time-scales.}
\label{fig:rin}
\end{figure}

\Cref{fig:rin} shows the output power of the \shg  on the long term (1 hour) and on the short term (\SI{2}{s}).
On the short term the amplitude noise is due to the residual frequency noise of the \hc locking.
When the cavity is properly locked the relative amplitude noise (root mean square) is below \SI{1}{\%} measured with a \SI{25}{kHz} bandwidth (\cref{fig:rin}).
On the long term the laser power shows good stability, with relative fluctuations of $\sim\SI{4}{\%}$ in two hours.

\begin{figure}[t]
\centering
\inputgnuplot{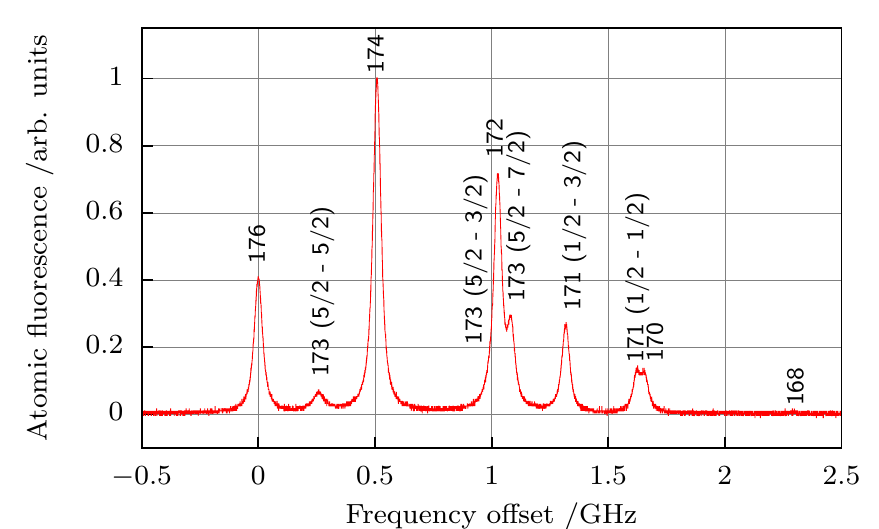}
\caption{Spectrum of ytterbium transition in a single sweep of the \tisa frequency. Isotope mass numbers  (hyperfine transitions for odd isotopes) label each resonance.}
\label{fig:ybspec}
\end{figure}

The laser is locked to the ytterbium \ce{^1S0 - ^1P1} transition by transverse spectroscopy on a thermal ytterbium beam to deal with the long term frequency drift.
The spectroscopy of the ytterbium transition is shown in \cref{fig:ybspec}, as the \tisa laser is scanned for \SI{1.5}{GHz}.
All stable isotopes of ytterbium are observed when the blue is tuned around a frequency of \SI{751.524}{THz}.
We have tuned the blue frequency in a \SI{280}{GHz} range (from \SI{751.36}{THz} to \SI{751.64}{THz}) without realignment of the optics and with a relative decrease in power of \SI{10}{\%} for the higher frequencies.
The source at \blue is  now used for a \mot, for a slower beam, and for a resonant probe beam in the \inrim optical frequency standard experiment \citep{Pizzocaro2013b}.

\section{Conclusions}
We have presented a reliable, powerful, single-frequency laser source at \blue based on the \shg in a \ac{lbo} crystal using an enhancement cavity.
A second harmonic power of \SI{1.0}{W} outside the cavity was achieved from \SI{1.22}{W} of infrared light.
We obtain a maximum efficiency of \SI{80}{\%} in good agreement with the expected value calculated from the measured crystal and cavity properties.
%The duplication works well and reliably, even if the input coupler is not exactely matched to the cavity losses.
%From numerical calculations we predict that replacing our input coupler with the optimal one results in a relative increase of the output power for $P\ped{in}=\SI{1.2}{W}$ below \SI{3}{\%}.

In atomic physics experiments with ytterbium atoms, the \blue source is a key component for the cooling, trapping, or atomic beam slowing on the strong \ce{^1S0 - ^1P1} transition.
This \blue source is well suited for this role being high-power and inheriting the narrow linewidth and tunability of the \tisa pump.
\ac{lbo} proved to be reliable at \blue even for high power.
The setup described here has been used without degradation of the crystal for one year at \inrim while a similar setup with the crystal under vacuum  and an output of \SI{0.6}{W} has been used for two years at \ac{ino} and \ac{lens}.
%The setup described here has been used without degradation of the crystal for 6 months and it is a key component for the loading of ytterbium in a lattice clock.
%A similar setup with the crystal under vacuum  and an output of \SI{0.6}{W} has been used for two years at \ac{ino} and \ac{lens}, where it has a similar important role to achieve high density ytterbium quantum gases.

\section{Acknowledgment}
The authors would like to acknowledge L. Fallani for careful reading, E. K. Bertacco for help with electronics and, A. Barbone, M. Bertinetti, and V. Fornero for technical help.
%The authors acknowledge for funding the European Metrology Research Programme (EMRP) Project SIB55-ITOC.
%The EMRP is jointly funded by the EMRP participating countries within EURAMET and the European Union.
The authors acknowledge for funding the European Metrology Research Programme (EMRP) Project SIB55-ITOC, the European Reserch Council (ERC) Advanced Grant DISQUA, the European Union 126 FP7 Integrated Project SIQS, and the European Union FP7 integrated infrastructure initiative Laserlab Europe.
The EMRP is jointly funded by the EMRP participating countries within EURAMET and the European Union.

%\bibliography{/home/marco/Documents/main}
\bibliographystyle{osajnl}
\bibliography{shg800-2}

\begin{thebibliography}{10}
\newcommand{\enquote}[1]{``#1''}

\bibitem{Hinkley2013}
N.~Hinkley, J.~A. Sherman, N.~B. Phillips, M.~Schioppo, N.~D. Lemke, K.~Beloy,
  M.~Pizzocaro, C.~W. Oates, and A.~D. Ludlow, \enquote{An atomic clock with
  $10^{-18}$ instability,} Science \textbf{341}, 1215--1218 (2013).

\bibitem{DeMille1995}
D.~DeMille, \enquote{Parity nonconservation in the
  $6{\mathit{s}}^{2}{}^{1}{\mathit{s}}_{0}\rightarrow{}6\mathit{s}5\mathit{d}{^{3}D}_{1}$
  transition in atomic ytterbium,} Phys. Rev. Lett. \textbf{74}, 4165--4168
  (1995).

\bibitem{Takasu2003}
Y.~Takasu, K.~Maki, K.~Komori, T.~Takano, K.~Honda, M.~Kumakura, T.~Yabuzaki,
  and Y.~Takahashi, \enquote{Spin-singlet bose-einstein condensation of
  two-electron atoms,} Phys. Rev. Lett. \textbf{91}, 040404 (2003).

\bibitem{Fukuhara2007}
T.~Fukuhara, Y.~Takasu, M.~Kumakura, and Y.~Takahashi, \enquote{Degenerate
  fermi gases of ytterbium,} Phys. Rev. Lett. \textbf{98}, 030401 (2007).

\bibitem{Pagano2014}
G.~Pagano, M.~Mancini, G.~Cappellini, P.~Lombardi, F.~Schafer, H.~Hu, X.-J.
  Liu, J.~Catani, C.~Sias, M.~Inguscio, and L.~Fallani, \enquote{A
  one-dimensional liquid of fermions with tunable spin,} Nat. Phys.
  \textbf{10}, 198--201 (2014).

\bibitem{Honda2002}
K.~Honda, Y.~Takasu, T.~Kuwamoto, M.~Kumakura, Y.~Takahashi, and T.~Yabuzaki,
  \enquote{Optical dipole force trapping of a fermion-boson mixture of
  ytterbium isotopes,} Phys. Rev. A \textbf{66}, 021401 (2002).

\bibitem{Hayes2007}
D.~Hayes, P.~S. Julienne, and I.~H. Deutsch, \enquote{Quantum logic via the
  exchange blockade in ultracold collisions,} Phys. Rev. Lett. \textbf{98},
  070501 (2007).

\bibitem{Park2003}
C.~Y. Park and T.~H. Yoon, \enquote{Efficient magneto-optical trapping of {Yb}
  atoms with a violet laser diode,} Phys. Rev. A \textbf{68}, 055401 (2003).

\bibitem{Adams1992}
C.~Adams and A.~Ferguson, \enquote{Tunable narrow linewidth ultra-violet light
  generation by frequency doubling of a ring {Ti}:sapphire laser using lithium
  tri-borate in an external enhancement cavity,} Opt. Commun. \textbf{90},
  89--94 (1992).

\bibitem{Kumagai2003}
H.~Kumagai, Y.~Asakawa, T.~Iwane, K.~Midorikawa, and M.~Obara,
  \enquote{Efficient frequency doubling of {1-W} continuous-wave {Ti}:sapphire
  laser with a robust high-finesse external cavity,} Appl. Opt. \textbf{42},
  1036--1039 (2003).

\bibitem{Koelemeij2005}
J.~C.~J. Koelemeij, W.~Hogervorst, and W.~Vassen, \enquote{High-power
  frequency-stabilized laser for laser cooling of metastable helium at 389 nm,}
  Rev. Sci. Inst. \textbf{76}, 033104 (2005).

\bibitem{Scheid2007}
M.~Scheid, F.~Markert, J.~Walz, J.~Wang, M.~Kirchner, and T.~W. H\"{a}nsch,
  \enquote{750 {mW} continuous-wave solid-state deep ultraviolet laser source
  at the 253.7 nm transition in mercury,} Opt. Lett. \textbf{32}, 955--957
  (2007).

\bibitem{Cha2010}
Y.-H. Cha, K.-H. Ko, G.~Lim, J.-M. Han, H.-M. Park, T.-S. Kim, and D.-Y. Jeong,
  \enquote{Generation of continuous-wave single-frequency {1.5 W 378 nm}
  radiation by frequency doubling of a {Ti}:sapphire laser,} Appl. Opt.
  \textbf{49}, 1666--1670 (2010).

\bibitem{Ferrari2010}
G.~Ferrari, J.~Catani, L.~Fallani, G.~Giusfredi, G.~Schettino, F.~Sch\"{a}fer,
  and P.~Cancio~Pastor, \enquote{Coherent addition of laser beams in resonant
  passive optical cavities,} Opt. Lett. \textbf{35}, 3105--3107 (2010).

\bibitem{Targat2005}
R.~L. Targat, J.-J. Zondy, and P.~Lemonde, \enquote{75\% efficiency blue
  generation from an intracavity {PPKTP} frequency doubler,} Optics
  Communications \textbf{247}, 471--481 (2005).

\bibitem{Boyd1968}
G.~D. Boyd and D.~A. Kleinman, \enquote{Parametric interaction of focused
  gaussian light beams,} J. Appl. Phys. \textbf{39}, 3597--3639 (1968).

\bibitem{Risk2003}
W.~P. Risk, T.~R. Gosnell, and A.~V. Nurmikko, \emph{Compact Blue-Green Lasers}
  (Cambridge University Press, New York, 2003).

\bibitem{Smith2009}
A.~V. Smith, \enquote{{SNLO} nonlinear optics code,}  (2009).
  {http://www.as-photonics.com/SNLO}.

\bibitem{Velsko1991}
S.~P. Velsko, M.~Webb, L.~Davis, and C.~Huang, \enquote{Phase-matched harmonic
  generation in lithium triborate ({LBO}),} IEEE J. Quantum Electron.
  \textbf{27}, 2182--2192 (1991).

\bibitem{Polzik1991}
E.~S. Polzik and H.~J. Kimble, \enquote{Frequency doubling with \ce{KNbO3} in
  an external cavity,} Opt. Lett. \textbf{16}, 1400--1402 (1991).

\bibitem{Freegarde2001}
T.~Freegarde and C.~Zimmermann, \enquote{On the design of enhancement cavities
  for second harmonic generation,} Opt. Commun. \textbf{199}, 435--446 (2001).

\bibitem{Hansch1980}
T.~H\"ansch and B.~Couillaud, \enquote{Laser frequency stabilization by
  polarization spectroscopy of a reflecting reference cavity,} Opt. Commun.
  \textbf{35}, 441--444 (1980).

\bibitem{Pizzocaro2013b}
M.~Pizzocaro, F.~Bregolin, D.~Calonico, G.~Costanzo, F.~Levi, and L.~Lorini,
  \enquote{Improved set-up for the ytterbium optical clock at {INRIM},} in
  \enquote{European Frequency and Time Forum International Frequency Control
  Symposium (EFTF/IFC), 2013 Joint,}  (2013), pp. 379--382.

\end{thebibliography}
%\bibliography{}

\end{document}